\documentstyle[12pt,aaspp4]{article}
\newcommand{\ms}{\mbox{m s$^{-1}~$}}
\newcommand{\msun}{M$_{\odot}~$}
\newcommand{\msune}{M$_{\odot}$}

\newcommand{\mjup}{M$_{\rm JUP}~$}
\newcommand{\mjupe}{M$_{\rm JUP}$}

\newcommand{\msini}{$M \sin i~$}

\lefthead{Fischer {\it et~al.\/}}
\righthead{Planetary Companions to {HD 195019 and HD 217107}}
\slugcomment{to appear in Jan 1999 Publications of the Astronomical Society of the Pacific}
\received{}
\accepted{}
\begin{document}

\title{Planetary Companions Around Two Solar Type 
Stars: HD 195019 and HD 217107$~^{1}$}

\author{Debra A. Fischer\altaffilmark{2},
Geoffrey W. Marcy\altaffilmark{2},
R. Paul Butler\altaffilmark{3}, 
Steven S. Vogt\altaffilmark{4}, 
Kevin Apps\altaffilmark{5}}

\authoremail{fischer@stars.sfsu.edu}

\altaffiltext{1}{Based on observations obtained at Lick Observatory, which
is operated by the University of California, and on observations obtained
at the W.M. Keck Observatory, which is operated jointly by the
University of California and the California Institute of Technology.}

\altaffiltext{2}{Department of Physics and Astronomy, San Francisco State University, 
San Francisco, CA, USA 94132
and at Department of Astronomy, University of California,
Berkeley, CA USA  94720, fischer@stars.sfsu.edu}

\altaffiltext{3}{Anglo--Australian Observatory, PO Box 296, NSW 1710 Epping, Australia}

\altaffiltext{4}{UCO/Lick Observatory, 
University of California at Santa Cruz, Santa Cruz, CA, 95064}

\altaffiltext{5}{Physics and Astronomy, Universityof Sussex, Falmer, Brighton, BN1, 9QJ, UK}

\begin{abstract}

We have enlarged the sample of stars in the planet search 
at Lick Observatory. Doppler measurements of 82 new 
stars observed at Lick Observatory, with additional velocities from  
Keck Observatory, have revealed two new planet candidates.  

The G3 V/IV star, 
HD 195019, exhibits Keplerian velocity 
variations with a period of 18.27 d, an orbital eccentricity of 
$0.03 \pm 0.03$, and \msini = 3.51 \mjupe.  Based on a measurement 
of Ca II H\&K emission, this star is chromospherically 
inactive.  We estimate the metallicity of HD 195019 
to be approximately solar from ubvy photometry.

The second planet candidate was detected around HD 217107, a G7V star.  
This star exhibits a 7.12 d Keplerian period with eccentricity $0.14 \pm 0.05$
and \msini = 1.27 \mjupe.  HD 217107 is also chromospherically inactive. 
The photometric metallicity is found to be [Fe/H]$ = +0.29 \pm 0.1$ dex.
Given the relatively short orbital period, the absence of tidal 
spin up in HD 217107 provides a theoretical constraint on the 
upper limit of the companion mass of $<11$ \mjupe.

\end{abstract}


\section{Introduction}
\label{intro}

In the past three years, a dozen extrasolar planet candidates have been 
discovered around main sequence stars 
using high precision radial velocity measurements.
All except one of these ``planetary'' companions has a derived mass 
(\msini) less than 5 times the mass of Jupiter.
Only three of the companions have orbital periods longer than one year.
One unexpected result has been that the orbits for 
the ``planets''  appear to be both circular and eccentric.  
The theoretical interpretation of this observation is that 
the planets formed in dissipative circumstellar disks, followed 
by gravitational perturbations (cf. Lin et al. 1996, Artymowicz 1997, 
Levison et al. 1998).  

One of the first of the extrasolar ``planets'', 51 Peg (Mayor \& Queloz 1995),
was a surprise: a 0.5 \mjup companion in a 4.2 day orbit.
With the latest discovery of a planet in a 3.1 day orbit 
around HD 187123 (Butler et al. 1998) and the companion to 
{$\tau$ Boo} and {$\upsilon$ And} (Butler et al. 1997) there are now 
four of these very short-period, 51 Peg - like systems.   
Models of tidal interactions between 
the star and planet (Terquem et al. 1998, Ford et al. 1998, Marcy et al. 1997) provide 
timescales for circularization and spin-up of the stellar 
convective envelope and constrain the companion mass 
to the planetary regime for cases where the orbital period is
$\sim 4$ days or less.  

A sample of 107 GK main sequence stars have been monitored  
since 1987 at Lick Observatory. A revision of the spectrograph optics 
in 1994 improved the internal radial velocity precision to  
3 m/s for the brighter stars in this program.  The Lick survey has 
identified ``planetary''
companions to 70 Vir, 47 UMa, 55 Rho Cnc, {$\tau$ Boo}, {$\upsilon$ And},
and GJ 876 (Marcy et al. 1998a) and confirmed the discovery of planets around 
51 Peg (Mayor \& Queloz 1995) and {$\rho$ CorBor} (Noyes et al. 1997).
A co-discovery of a companion to 16 Cyg B (Cochran et al. 1997) was 
also made with observations from Lick Observatory.   
The long term nature of 
the Lick Observatory program and the high precision of the 
doppler analysis should continue to provide information 
regarding low-amplitude, longer-period companions. 

In the past few months, 200 new stars have been added to 
the Lick project.  The first goal in extending this planet-search 
program is to identify short-period 
companions to help characterize the planetary mass function and the 
orbits of such systems. These short-period systems test 
theories of tidal interactions between the host star and 
its companion and provide ideal targets for transit observations.   
An observation of a planet-transit would provide powerful corroboration
regarding the nature of the planet candidates.

\section{The Second Generation Lick Project}
\label{lick}

The selection criteria for the new sample stars were dictated by the fact 
that high S/N and an abundance of narrow spectral lines is 
required to achieve good velocity precision.  Stars brighter 
than V=7.5 with F7 - K0 spectral type were selected 
from the Hipparcos catalog and their main sequence status 
was confirmed using Hipparcos parallaxes to derive the absolute 
visual magnitude.  This yielded a preliminary list of stars 
that were not already being observed in the Keck or Lick planet search programs. 

The precise Doppler technique makes use of an iodine cell to 
impose a grid of sharp reference lines on the stellar spectrum.
Also needed is a ``template'' observation of every sample star 
without iodine to serve as a reference.  The Doppler analysis code 
later convolves this template spectrum with an FTS iodine observation 
to model subsequent observations of the star with iodine and thereby
derive the spectrometer PSF and calibrate the 
wavelength (velocity) shift (Butler et al. 1996).  
High quality template observations for the sample stars 
were obtained with HIRES (Vogt 1994) at the Keck Observatory with 
R=87,000 and S/N of 300. 

The Keck templates were immediately used to estimate vsini, 
measure Ca II H\&K emission and check
for double lines in the spectra.  High vsini reduces the velocity precision 
that can ultimately be obtained and a double-line spectrum 
precludes the Doppler analysis entirely. Core emission 
in the Ca II H\&K lines is characterized by a pseudo-equivalent 
width, or S index, analogous to the Mount Wilson 
HK observations (Baliunas 1998, Shirts \& Marcy 1998). The S index 
is then transformed to R'HK, a ratio of the HK flux to the bolometric 
flux of the star. R'HK is a good indicator of the level of chromospheric 
activity and a good 
predictor of the rotational period and the age of main 
sequence stars (Noyes et al. 1984).
Strong chromospheric activity is correlated with magnetic 
activity and motions in the
photosphere of the star which can produce quasi-periodic radial 
velocity variations (Saar et al. 1998, Marcy \& Butler 1998, 
Butler et al. 1998), so 
low amplitude velocity variations in active stars warrant caution. 

After evaluating the template spectra, we ended up with 
82 new sample stars that were accessible in July and August from 
Lick Observatory. 
In order to search all of these stars 
with a minimum expenditure of telescope time, we opted to 
take 2-5 minute exposures which yielded S/N $\sim 100$ with 
typical velocity precisions of only 10 -- 15 m/s. In 
retrospect, we believe this lower precision and limited sampling 
biased us against finding low amplitude objects like 51 Peg.  
We intend to continue monitoring 
these objects long-term and with higher precision in the future to 
search for longer-period and lower-amplitude companions.  

We used the 3m Shane telescope with the Hamilton spectrograph (Vogt 1987) 
at Lick Observatory to obtain 4-5 observations of each of the new sample 
stars on July 9-11, July 30,31 and Aug 1, 1998.  
Velocity dispersions greater than $3 \sigma$ were found in nine stars.  
A timely allocation of Keck telescope time made it possible 
to add these nine stars to the Keck planet search program and
obtain high S/N spectra ($\sim 300$) during 7 of 8 consecutive nights 
from Sept 12-19.  The Keck observations showed that six of the 
nine candidates were not varying in short-period orbits, however 
three of the candidates did show Keplerian-like velocity shifts. 
We then used the 0.6 meter Coude Auxilliary Telescope (CAT) 
at Lick Observatory 
to monitor these three stars on a nightly basis.
Typically, a set of three half-hour CAT exposures were obtained within a 2 hour 
interval.  Velocities determined within a two hour interval were averaged 
to form a single velocity measurement. 

The Lick and Keck velocities each have 
independent, arbitrary velocity
zero--points.  The relative zero--point 
was determined by combining the two data sets and
adjusting the velocity offset until the Keplerian
fit yielded a minimum in the RMS velocities. This exercise forced  
the Lick and Keck velocities to have the same zero-point.
After the Lick and Keck observations were combined, the 
velocity variations for two of the three candidates (HD 195019 and 
HD 217107) phased up nicely.  
In the case of the third candidate, HD 187897,
the velocities did not phase up.  Subsequent 
CAT observations revealed a departure from Keplerian velocity 
in the Lick data alone.  
The presence of moderate core emission
in the Ca II H\&K lines 
had raised suspicions regarding 
this star (Figure 1 - bottom plot). 
For HD 
187897, we derive an S index of 0.286 and $\log R'HK=-4.5$.   
This level of chromospheric activity has been shown (Saar \& Donahue 1997) 
to produce radial velocity variations with amplitudes 
between 30 - 50 m/s, and we therefore believe that the 36 m/s semi-amplitude 
Doppler shifts observed in HD 187897 may 
originate from activity-related photospheric variations in the star.

\section{Observations: HD 195019}
\label{obs}

HD 195019 (=HIP 100970) is spectral type G3 V/IV 
with V=6.87, B-V = 0.662 from Hipparcos photometry (Perryman et al. 1997).  
Based on the spectral type, we assume a mass 
of $0.98 \pm 0.06$ \msun (Lang 1992). 
This star has a companion, ADS 13886B, that is fainter by 3 magnitudes 
and separated by about 4 arcseconds.  The linear separation 
of the two components is about 150 AU so this stellar companion has 
no impact on the velocities discussed here.  The spectrum of HD 195019 
is not contaminated by AC 13886B because this secondary star is 
relatively faint and the angular separation 
is sufficient to spatially resolve the two components.

Figure 1 (top plot) shows a spectral 
window centered on the Ca II H line for HD 195019.
There is no apparent Ca II H\&K core emission 
and the derived S value (0.189) implies a rotation period of 22 d, 
$\log R'HK=-4.85$ and $\log age=9.5$ years. 
The spectral lines are narrow, supporting slow stellar 
rotation and we measure a radial velocity of $-70.3 \pm 2$ km/s. 
The space motions for this star are consistent with 
those of old disk stars.  The Hipparcos parallax of 26.77 mas 
yields $M_V=4.01$, about 0.7 magnitudes brighter than 
expected for a ZAMS G3 star.  There are 
two possible explanations which we cannot distinguish 
between at this time.  If the Ca II H\&K diagnostic is 
correct then the star could 
be metal-rich.  We estimate that [Fe/H] $\sim +0.30$ could 
result in the observed shift from the ZAMS for this star.  If HD 195019 
is metal rich, it may be akin to $\tau$ Boo 
which has [Fe/H] = +0.25 (Gonzalez 1998). However,  
this interpretation would require significant
contamination of the ubvy photometry by the unresolved companion 
and seems unlikely given the relative faintness of the companion.
Alternatively, the star may be slightly evolved with solar 
metallicity (from ubvy photometry), consistent with the 
luminosity classification and observed space motions.  
In either case, the chromospheric activity 
in this star is low and we expect that activity-associated spots 
(cf, Butler et al. 1998) are an unlikely explanation for the 
observed Doppler shifts.  Photometric observations of HD 195019 would 
provide additional discriminating evidence. 

 The spectral format at Lick Observatory includes the 
Li I resonance line at $\lambda 6707.8 \, \AA$. The spectral 
resolution of the Hamilton spectrograph, R= 50,000, 
is sufficient to resolve the Li I line from an 
adjacent feature at $\lambda 6707.45 \, \AA$.  Three consecutive 
observations were coadded to build the S/N to 140. 
The composite spectrum was then used to derive 
$\log {\rm N(Li)} \le 0.6$ in HD 195019. This lithium abundance was 
determined from a bilinear interpolation of a lithium curve of 
growth (Soderblom et al. 1993a) and is given 
on a scale where $\log {\rm N(H) = 12.0}$. The abundance determination 
assumes $T_{eff}$=5600K (derived from a calibration 
to B-V=0.66, Soderblom et al. 1993b) and an upper limit to the line 
equivalent width of $2.0\,{\rm m}\AA$. Since the B-V color is 
uncorrected for the unresolved stellar companion, the true temperature
of the star may be hotter by almost 100K. This would increase the 
uppper limit in the derived lithium abundance to 0.8.  For comparison, the 
abundance of lithium in the sun is $\log {\rm N(Li)} \sim 1.0$.

Twelve velocity measurements for HD 195019 were obtained over a 
4-month period at Lick Observatory and combined with 
seven observations from Keck 
as previously described.  To phase the Lick and Keck data, a velocity 
shift of -128 m/s was applied to all of the Keck velocities.  The velocities
(shifted to a common zero point), 
Julian date of observations and the telescope used to obtain the velocites 
are listed in Table 1. In Table 1, ``Shane'' is the 3m Shane telescope 
at Lick Observatory, ``Keck'' is the 10m Keck telescope and 
``CAT'' is the 0.6 Coude Auxilliary Telescope at Lick.  
A periodogram analysis reveals a strong 
peak at 18.27 days for the combined data set. 
The best-fit Keplerian is shown in 
Figure 2. The internal velocity errors from the S/N$\sim 100$ observations 
with the 3m Shane telescope at Lick were initially 
about 12.8 m/s.  These were reduced to about 7 m/s by 
combining velocity measurements from three consecutive 
half hour CAT observations after 
this star was identified as having a planet candidate.  The Keck observations 
had S/N $\sim 300$ with velocity errors of about 6 m/s.  
The internal velocity errors from Keck 
stem both from photon statistics (2 m/s) and systematic errors which 
are currently being investigated.  Figure 3 shows 
a phased velocity plot containing all data for HD 195019. 
The triangles represent Keck 
observations and the circles represent 
Lick observations. 
The best-fit Keplerian for the combined data yields 
an orbital period of $18.27 \pm 0.14$ days, a semi-amplitude 
K=$275 \pm 5$ m/s, and 
an eccentricity of $0.03 \pm 0.03$.
The RMS to the fit of the Keplerian curve is 12.8 m/s. 
With our assumed mass of 0.98 \msun  for HD 195019,  
the orbital solution suggests a 
companion mass \msini = $3.51 \pm 0.4$ \mjupe. The uncertainty in 
the mass of the companion is derived from the uncertainty 
in the stellar mass.  The orbital elements for 
HD 195019 are summarized in Table 2.

\section{Observations: HD 217107}
\label{obs}

The Hipparcos catalog lists a spectral classification of G8IV for 
HD 217107 (=HR8734, = HIP113421) and B-V=0.744.  
The apparent magnitude, V=6.17, and Hipparcos parallax of 50.71 mas provide 
$M_V = 4.70$, about 0.7 mag above the zero age main sequence (Lang 1992). 
Our calibration of ubvy photometry suggests that this is a 
G7V star with a metallicity [Fe/H]=$0.29 \pm 0.1$. 
Fine spectral analysis should be carried out for this star to 
verify the metallicity derived here. This higher 
metallicity would explain the position 
of this star on the CMD. For a solar metallicity G7 or G8 star, 
the mass estimate would 
be approximately 0.9 \msune. However, we estimate that 
the enhanced metallicity 
will result in a correction to the mass of the star of 0.1 \msune.  
We adopt an intermediate value of $0.96 \pm 0.06$ 
\msun for HD 217107. 
The lack of emission in the Ca II H\&K lines (Figure 1 - middle plot) 
implies that this 
star is chromospherically inactive.  
The measured S value (0.15) yields Prot=39d, $\log R'HK=-5.0$, 
log age=9.89 years.  
This star has disk motions similar to the Sun with an unusually low 
transverse velocity ($<2$km/s) and a radial velocity of -12.1 km/s. 

The lithium abundance is also subsolar in this star. Three consecutive 
observations with composite S/N of 180 were used to set an upper
limit in the equivalent width of Li I $ \lambda 6707.8 \, \AA$ 
as $2 {\rm m} \AA$.  Adopting $T_{eff}=5360$ (calibrated to B-V=0.744), 
we find $\log {\rm N(Li)} \le 0.4$.

Fourteen velocity measurements were obtained at Lick Observatory over the
4-month interval from July - October 1998, and were combined 
with seven observations from Keck.
The internal velocity errors at Lick were initially 
9.6 m/s and were later reduced to about 6 m/s by 
combining velocity measurements for three half hour CAT observations
obtained after HD 217107 was confirmed as a 
planet candidate at Keck. The internal 
velocity errors at Keck were about 6 m/s. The velocity offset between 
Keck and Lick data was found to be zero and the velocities are listed in 
Table 3. A periodogram analysis of the combined velocities revealed 
a strong peak at 7.11 days.  A Keplerian fit to the data,
shown in Figure 4, yields 
the following orbital parameters: an orbital 
period of 7.12$ \pm 0.02$ d, K=139.5$ \pm 4.1$ m/s, 
eccentricity=0.14$ \pm 0.05$ (see Table 2). The Keplerian fit had an 
RMS=14.6 m/s.
The inferred companion mass, \msini, is $1.27 \pm 0.4$ \mjupe.
The phased plot for the combined data sets is shown in Figure 5. 


\section{Discussion}

We have added nearly 200 new stars to the Lick planet search. 
In July and August, we observed 82 of these stars 
and here we present two planet candidates around the  
chromospherically-quiet G dwarfs, HD 195019 and HD 217107.
The planet around HD 195019 has an orbital period of 18.27 days, 
similar to 55 Cancri (P=14.7 d). The periodicity in the radial 
velocities is similar to the rotation period (22 days) derived 
from a measure of the Ca II H\&K emission.  
However, while astrophysical 
processes can mimic Keplerian velocity signals in chromospherically 
active stars with periodicities of a few days, 
it is not clear that modulation would occur in a chromospherically 
inactive star like HD 195019.  Photometric observations will provide
an important diagnostic here.  The orbital eccentricity of HD 195019 
is $0.03 \pm 0.03$ and the companion mass, \msini, is 3.51 \mjupe.  
The companion to HD 217107 has 
an orbital period of 7.12 days, an eccentricity 
of $0.14 \pm 0.05$ and \msini = 1.27 \mjupe.  
Tidal effects are not expected to have circularized 
gas giants with orbital periods longer than a few days 
(Terquem et al. 1998, Ford et al. 1998, Marcy et al. 1997), so 
the nearly circular orbits derived for these two stars may be primordial.  

Based on ubvy photometry, HD 195019 appears to have solar 
metallicity and HD 217107 appears 
to be metal rich with [Fe/H]=+0.29. This lends some additional evidence 
favoring the suggestion that stars with close-in, gas giants may be metal 
rich relative to the general population of field stars (Gonzalez, 1998).
Fine spectral analysis of both stars is needed to confirm these results. 

Figure 6 shows the distribution of detected planetary 
companions in a two-parameter plane of companion mass (in \mjupe) and 
orbital period. 
The points representing HD 195019 and HD 217107 are labeled
so that they may be more easily compared with the other 
extrasolar planet detections.  Figure 6 also includes two 
planet detections, HD 168443 and HD 210277 
(Marcy et al. 1998b), 
slightly in advance of publication because they add  
to the interpretation of this figure.  Both of these soon-to-be announced
planets have highly eccentric orbits that further highlight the 
observation that eccentric orbits are not uncommon. 
Orbits with eccentricities exceeding 0.20 are circled with 
an ellipse.  The choice of a 0.2 eccentricity threshold is 
arbitrary; it was chosen because it marks cases where the 
derived eccentricity clearly exceeds the uncertainty. 

Also shown in Figure 6 are two radial velocity detectability 
curves. The detectability curves are based on Monte Carlo 
simulations that add gaussian noise to the velocity curve 
of a test secondary mass with a given orbital period. For 
orbital periods longer than the extent of the radial velocity 
program, each phase of the velocity curve is tested.  When 
the scatter in the velocity plus noise is detectable at the 
95\% confidence level by an F-ratio test, the secondary mass 
is deemed detectable.  
One of the curves demarkates the (small) regime 
in this two-parameter space in which planets could have been  
detected during our initial observations of new sample stars 
last July. It assumes an observing interval of one month
with 5 evenly spaced observations 
per star, circular orbits, an average value for the inclination ($\pi / 4$),  
and a velocity precision of 10 m/s.  
The second detectability curve makes the same assumptions but with 
an observing interval of 8 years and 4 observations per star 
per year (typical of the original, ongoing Lick program).  Given the 
distribution of detected companions in Figure 6, even the limited 
sampling of 5 observations over a one month interval 
appears to be fairly efficient sieve for planets in the new 
sample. However, analogs to 51 Peg with low velocity amplitudes 
could have escaped detection with the velocity precision and 
sampling frequency employed here.  
It is interesting that the parameter 
space outside of the ``one month'' detectability curve is not 
particularly rich in companions. 
Such companions could have been detected in the long 
term project at Lick Observatory.  The absence of brown dwarfs 
which would have been easily detected is also striking. 

We derive \msini, rather than $M$, for the planet candidates. 
The question of whether 
these planets could actually be brown dwarfs or low mass stars 
has been considered by Marcy and Butler (1996).  Spectroscopic 
companions with \msini = 5 \mjup   
would have stellar masses (above 70 \mjup) 
if the orbit were 4 degrees or less from 
a face-on orientation. Statistically, this inclination will occur in 
about 0.2\% of randomly-oriented orbits.  To estimate the probability 
that a planet candidate is actually a stellar companion, 
the 0.2\% inclination probability
must be multiplied by the probability that a star 
is a binary system with a low mass companion in the relevant 
separation range. Integrating the mass and separation distribution 
(Duquennoy \& Mayor 1991) suggests that about 
8\% of solar-type stars could have such a stellar companion. 
So, statistically, 1 in 6,250 surveyed stars could disguise 
a stellar companion as a 5 \mjup  spectroscopic companion, but only 
a few hundred solar type stars have been surveyed for these companions. 
Alternatively, a 5 \mjup  companion be a 40 \mjup brown dwarf. 
Since the mass function of brown dwarfs 
is unknown, we can only set an upper limit to this probability based 
on the low orbital inclination probability.  However, 
there appears to be a vanishingly small number of companions with 
\msini greater than 10 \mjupe, suggesting that brown dwarf companions
to solar type stars are rare (Mayor et al. 1999). 

Additional constraints supporting the low mass, planetary nature of these 
companions comes from models of tidal interactions between the star and 
planet. 
If the companion to HD 217107 were more 
massive than 11 \mjupe, it is expected that tidal torques would have 
spun up the convective envelope of the star to match the orbital period 
(Terquem et al. 1998).  However the rotation period of the star is 
estimated to be 39 days (based on the Ca II H\&K lines) while 
the orbital period is only 7.11 days. 

The survey at Observatoire de Haute-Provence (Mayor \& Queloz 1995), 
discovered 51 Peg in a survey of 140 solar type stars.  
The original Lick project surveyed 107 stars, and 1.9\% of those 
stars ($\tau$ Boo, $\upsilon$ And; Butler et al. 1997) were found to have 
companions with circularized orbits and periods of a few days. 
The 82 new Lick stars were 
observed 4-5 times over a 3 week interval with an initial precision 
of only 10 m/s in the hopes of identifying 51 Peg analogs. 
Neither of the planets presented in this paper 
qualify as 51 Peg analogs as their orbital periods are too long.  
It is possible that there 
are no such systems in the sample that 
we have observed so far.  However we do not believe this is 
a conclusive result. In retrospect, we believe that the initial lower 
precision of 10 m/s and the small number of observations (4 or 5 per star) 
was inadequate to detect the one or two low-amplitude, 
short-period systems that might still exist in the sample.  
We will continue to monitor 
these new additions to the Lick program long-term 
and with higher Doppler precision.

\acknowledgements

We thank Phil Shirts for the measurement of Ca II H\&K in 
the new sample templates.  Mike Eiklenborg obtained  
some of the observations. 
We acknowledge support by NASA grant
NAGW-3182 and NSF grant AST95-20443 (to GWM), and by NSF grant
AST-9619418 and NASA grant NAG5-4445 (to SSV) and by Sun Microsystems.
We thank the NASA and UC Telescope assignment committees for
allocations of telescope time.

\clearpage
\begin{figure}
\figcaption{The spectral window centered on the Ca II H line at $\lambda 3970\, \AA$
for HD 195019 (top), HD 217107 (middle) and HD 187897 (bottom). 
Core emission, indicating chromospheric activity, can be seen in HD 187897.}

\label{fig1}
\end{figure}

\begin{figure}
\figcaption{The combined Lick and Keck radial velocities for HD 195019.
The solid line is the radial velocity curve from the best--fit 
orbital solution.}
\label{fig2}
\end{figure}

\begin{figure}
\figcaption{The combined Lick and Keck radial velocities plotted vs 
orbital phase for 
HD 195019.  The filled circles represent Lick observations and the 
triangles represent Keck observations. 
The solid line is the radial velocity curve from the orbital solution
for the combined data set.}

\label{fig3}
\end{figure}

\begin{figure}
\figcaption{The combined Lick and Keck radial velocities for 
HD 217107.  
The solid line is the radial velocity curve from the orbital solution
for the combined data set.}

\label{fig4}
\end{figure}

\begin{figure}
\figcaption{The combined Lick and Keck radial velocities for HD 217107,
plotted versus orbital phase.  Filled circles are from Lick and triangles
come from Keck.
The solid line is the radial velocity curve from the orbital solution.}
\label{fig5}
\end{figure}

\begin{figure}
\figcaption{The detected planet companions from all surveys to date, including
two soon-to-be announced planets (Marcy et al. 1998b).  
The planets in this paper, 
HD 195019 and HD 217107, are labeled.  Orbital solutions with eccentricities 
greater than 0.20 are circled with ellipses.  Two radial velocity 
detectability curves are plotted.  The curve rising up 
to the left of the plot simulates a one month observing 
interval with 5 observations per star (similar to our campaign in 
July).  Planets with orbtital periods to the left of this 
curve could have been detected.  The other curve models the 
detectability regime of the standard Lick planet search. It 
assumes an observing interval of 8 years with 4 observations per year. 
Companions above this longer detectability curve could have 
been found in the sample of stars observed at Lick since 1987. 
Both curves assume a radial velocity precision of 10 m/s. }
\label{fig6}
\end{figure}

\begin{table}
\tablenum{1}
\caption{Radial Velocities for HD 195019}
\begin{tabular}{crrrrrr}
\tableline
\tableline
            JD    &   RV          & Telescope   \nl
       -2450000   &  (m/s)        &           \nl
\tableline
      1004.874  &        270.61  & Shane  \nl 
      1006.883  &        146.82  & Shane  \nl 
      1026.844  &        -11.26  & Shane  \nl 
      1027.869  &        -70.12  & Shane  \nl 
      1045.836  &        -34.42  & Shane  \nl 
      1068.852  &       -222.53  & Keck   \nl 
      1069.893  &       -185.77  & Keck   \nl 
      1070.913  &       -106.85  & Keck   \nl 
      1071.849  &          0.00  & Keck   \nl 
      1072.837  &         92.42  & Keck   \nl 
      1074.857  &        244.19  & Keck   \nl 
      1075.795  &        292.32  & Keck   \nl 
      1076.733  &        325.68  & Keck   \nl 
      1077.802  &        262.04  & CAT    \nl 
      1078.748  &        257.65  & CAT   \nl 
      1079.786  &        174.90  & CAT   \nl 
      1081.675  &         -3.40  & CAT   \nl 
      1082.669  &        -79.72  & CAT   \nl 
      1101.719  &       -150.00  & CAT   \nl 
\tableline
\end{tabular}

\end{table}

\begin{table}
\tablenum{2}
\caption{Orbital Parameters}
\begin{tabular}{crrrrrrrrrrr}
\tableline
\tableline
Param          		& HD 195019             & HD 217107 \nl
\tableline
P  (d)         		& 18.27 (0.14)            & 7.12 (0.02)     \nl
$^1{\rm T}_{\rm max}$ (JD)    & 2451072.16 (1.6)      & 2451067.42 (0.30) \nl
e              		& 0.03 (0.02)           & 0.14 (0.05)      \nl
$^2\omega$ (deg) 	& 250 (31)              & 19 (14)          \nl
K$_1$ (\ms)    		& 275.28 (5.0)          & 139.5 (4.1)      \nl
a$_1 \sin i$ (AU)	& $4.6 \times 10^{-4}$  & $9.0 \times 10^{-5} $  \nl
f$_1$(m) (M$_\odot$)  & $3.94 \times 10^{-8}$ & $1.94 \times 10^{-9}$ \nl
M$_2 \sin i$ (M$_{Jup}$) & 3.51 (0.4)           & 1.27 (0.4)       \nl                  
$^3{\rm Nobs}$          & 19                    & 21               \nl
\tableline
\end{tabular}

\tablenotetext{}{$^1$ Time of velocity maximum.}
\tablenotetext{}{$^2$ $\omega$ is poorly constrained for nearly circular orbits.}
\tablenotetext{}{$^3$ Number of observations from both Lick and Keck. Velocities 
obtained within a 2 hour period at Lick are averaged values.}
\end{table}

\clearpage

\begin{table}
\tablenum{3}
\caption{Radial Velocities for HD 217107 }
\begin{tabular}{crrrrrr}
\tableline
\tableline
            JD    &  RV          & Telescope   \nl
      -2450000    & (m/s)        &             \nl
\tableline
       1005.962  &      -76.55   & Shane  \nl
       1006.967  &      -68.69   & Shane  \nl
       1014.918  &      -29.73   & Shane  \nl
       1025.980  &       39.86   & Shane  \nl
       1027.941  &      -61.34   & Shane  \nl
       1049.811  &      -73.70   & Shane  \nl
       1068.860  &       54.01   & Keck   \nl
       1069.973  &      -41.24   & Keck   \nl
       1070.954  &      -49.35   & Keck   \nl
       1071.869  &       -6.99   & Keck   \nl
       1072.929  &      101.97   & Keck   \nl
       1074.870  &      195.79   & Keck   \nl
       1075.835  &       71.13   & Keck   \nl
       1076.800  &      -35.83   & Keck   \nl
       1077.872  &      -56.74   & CAT    \nl
       1078.817  &        0.00   & CAT    \nl
       1079.724  &       71.96   & CAT    \nl
       1079.859  &       92.17   & CAT    \nl
       1081.743  &      225.55   & CAT    \nl
       1100.756  &       35.90   & CAT    \nl
       1101.759  &      127.44   & CAT    \nl
\tableline
\end{tabular}

\end{table}


\clearpage

\begin{thebibliography}{}
\parsep 0pt
\itemsep -3pt

\bibitem[Artymowicz 1997]{artym:97}
Artymowicz P. 1997, in ASP Conf. Proc. 134, Brown Dwarfs and
Extrasolar Planets: Proceedings of a Workshop held in Tenerife, Spain,
17 - 21 March 1997, ed. R. Rebolo, E.\ L. Mart\'\i n, \& 
M.\ R. Zapatero Osorio (San Francisco: ASP), 152

\bibitem[Baliunas {  et~al.} 1998]{Bal98}
Baliunas, S.~L., Donahue, R.~A., Soon, W., \& Henry, G.\ W. 1998,
in ASP Conf. Proc 154, The Tenth Cambridge Workshop on Cool Stars, 
Stellar Systems and the Sun, ed. R.\ A. Donahue and J.\ A.
Bookbinder (San Francisco: ASP), 153 




\bibitem[Butler { et~al.} 1996]{BuMaWi96} Butler, R.~P., Marcy, G.~W.,
Williams, E., McCarthy, C., Dosanjh, P., \& Vogt, S.~S.  1996,
\newblock { PASP, } {108}, 500

\bibitem[Butler {  et~al.} 1997]{BuMaWi97}
Butler, R.~P., Marcy, G.~W., Williams, E., Hauser, H., \& Shirts, P.  1997,
\newblock { ApJ, } {474}, L115

\bibitem[Butler { et~al.} 1998]{BuMaVo98} Butler, R.~P., Marcy, G.~W.,
Vogt, S.~S., \& Apps, K.  1998,
\newblock { PASP, } in press

\bibitem[Cochran {  et~al.} 1997]{CoHa97}
Cochran, W.~D., Hatzes, A.~P., Butler, R.~P., \& Marcy, G.~W.  1997,
\newblock {  ApJ, } {483}, 457


\bibitem[Duquennoy \& Mayor 1991]{DM91}
Duquennoy, A., Mayor, M.  1991, 
\newblock { A\&A, } 248, 485

\bibitem[Ford et al. 1999]{ford:98}
Ford, E.\ B., Rasio, F.\ A., \& Sills, A. 1998, 
\newblock { ApJ,} {preprint}


\bibitem[Gonzalez 1998]{gonz:98}
Gonzalez, G. 1998, 
\newblock { A\&A, } 334, 221





\bibitem[Lang 1992]{lang:92}
Lang, K.\ R. 1992, Astrophysical Data: Planets and Stars.
(2nd ed.; New York, NY: Springer-Verlag), 132 

\bibitem[Levison {et al.} 1998]{lld:98}
Levinson, H.\ F., Lissauer, J.\ J., \& Duncan, M.\ J. 1998,
\newblock { AJ, } in press

\bibitem[Lin {  et~al.} 1996]{LiBoRi96}
Lin, D. N.~C., Bodenheimer, P., \& Richardson, D.~C.  1996,
\newblock {  Nature, } {380}, 606




\bibitem[Marcy \& Butler 1996]{MaBu96}
Marcy, G.~W. \& Butler, R.~P.  1996,
\newblock {  ApJ, } { 464}, L147

\bibitem[Marcy \& Butler 1998]{ann96}
Marcy, G.~W. \& Butler, R.~P.  1998,
\newblock {  ARA\&A, } 36, 57

\bibitem[Marcy { et~al.} 1997]{MaBuWi97} Marcy, G.~W., Butler, R.~P.,
Williams, E., Bildsten, L., Graham, J.~R., Ghez, A., \& Jernigan, G.
1997, \newblock { ApJ}, 481, 926

\bibitem[Marcy { et~al.} 1998]{MBVFA97} Marcy, G.~W., Butler, R.~P.,
Vogt, S.\ S., Fischer, D.\ A. \& Apps, K. 
1998a, \newblock { ApJ, } 505, L147 

\bibitem[Marcy { et~al.} 1998]{MBVL98} Marcy, G.~W., Butler, R.~P.,
Vogt, S.\ S., Fischer, D. A., \& Liu, M. 1998b, preprint  

\bibitem[Mayor \& Queloz 1995]{MaQu95}
Mayor, M. \& Queloz, D.  1995,
\newblock {  Nature, } {378}, 355


\bibitem[Mayor {  et~al.} 1999]{MaQuUd99}
Mayor, M., Queloz, D., Udry, S., \& Halbwachs, J.-L.  1999,
to appear in IAU Colloq. 170, Precise Radial Velocities


\bibitem[Noyes et al. 1997]{noyes:97} Noyes, R.\ W., Hartmann, L.\ W.,
Baliunas,S.\ L., Duncan, D.\ K., \& Vaughan, A.\ H. 1984, { ApJ} 279, 763

\bibitem[Noyes et al. 1984]{noyes:84} Noyes, R.\ W., Jha, S.,
Korzennik,S.\ G., Krockenberger, M., Nisenson, P., Brown, T.\ M.,
Kennelly, E.\ J., \& Horner, S.\ D. 1997, { ApJ,} 483, L111


\bibitem[Perryman et al. 1997]{Perry97}
Perryman, M. A. C., et al. 1997, A\&A, 323, L49 



\bibitem[Saar {et al.} 1998]{saar:98} Saar, S.\ H., Butler, R.\ P.,
\& Marcy, G.\ W. 1998, { ApJ,} 498, L153

\bibitem[Saar \& Donahue]{sd:97} Saar, S.\ H., \& Donahue, R.\ A. 1997, 
{ ApJ} 485, 319 


\bibitem[Shirts \& Marcy 1998]{sm:98} Shirts, P., \& Marcy, G.\ W. 1998, 
preprint 

\bibitem[Soderblom {et~al.} 1993]{so:93}
Soderblom,  D. R., Jones, B. F., Balachandran, S., 
Stauffer, J. R., Duncan, D. K., Fedele, S. B., \& Hudon, J. D. 1993,
\newblock{ AJ,} 106, 1059

\bibitem[Soderblom {et~al.} 1993]{ss:93}
Soderblom, D. R., Stauffer, J. R., Hudon, J. D., \& Jones, B. F. 1993,
\newblock{ ApJS, } 85, 313

\bibitem[Terquem {et al.} 1998]{tpnl:97} Terquem, C., 
Papaloizou, J.\ C.\ B., Nelson, R.\ P., \& Lin, D.\ N.\ C. 1998, 
{ ApJ,} 502, 788



\bibitem[Vogt 1987]{vogt:87}
Vogt, S. S. 1987, PASP, 99, 1214

\bibitem[Vogt et al. 1994]{vogt:94}
Vogt S. S. et al. 1994, Proc. SPIE, 2198, 362




\end{thebibliography}
\end{document}